\documentclass[a4paper,11pt]{article}
\usepackage{pos}
\usepackage{float}
\usepackage{subcaption}
\usepackage{lineno}
\usepackage{comment}
\usepackage{multirow}

\renewcommand{\printHeadAuthors}{T. Capistrán et al.}

\title{Machine learning reconstruction of cosmic ray parameters in EAS at HAWC}
\ShortTitle{ML reconstruction of CR parameters in EAS at HAWC}

\author[a]{J. Jaimes}
\author*[b]{T. Capistrán}
\author[c]{I. Torres}
\onbehalf{on behalf of the HAWC Collaboration}

\affiliation[a]{Universidad Industrial de Santander,
Calle 9 \#27, Bucaramanga, Colombia}

\affiliation[b]{Università degli Studi di Torino,
I-10125 Torino, Torino, Italy}

\affiliation[c]{Instituto Nacional de Astrofísica, Óptica y Electrónica, Luis Enrique Erro \#1, Puebla, Mexico}

\emailAdd{jjalfredo68@gmail.com}
\emailAdd{tomas.capistranrojas@unito.it}
\emailAdd{ibrahim@inaoep.mx}

\abstract{The High-Altitude Water Cherenkov (HAWC) Observatory comprises 300 water Cherenkov detectors, each equipped with four photomultipliers, located on the Volcán Sierra Negra in Mexico at 4,100 masl. This observatory can detect gamma rays in an energy range from 300~GeV to 100~TeV and cosmic rays from 100~GeV to 1~PeV. One of HAWC’s primary challenges is characterizing air showers and estimate their physical parameters, a highly complex task due to the nature of the data and the processes involved. Currently, HAWC employs two energy estimators for gamma rays: the ground parameter method and a neural network-based approach. However, for cosmic rays, only the likelihood-based estimator is available. In this work, we leverage machine learning techniques to achieve more accurate estimation of the physical parameters of cosmic rays. These techniques are explored as an alternative for reconstructing the physical properties of extensive air showers using simulated data aligned with the observatory’s configuration. Various models were trained and evaluated through an optimized pipeline and the most effective one was selected as the final implementation after a comprehensive comparison. This approach improves the accuracy of physical parameter estimation, contributing significantly to the detailed characterization of cosmic ray events.}

\ConferenceLogo{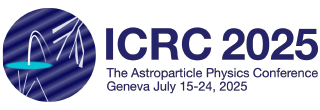}

\FullConference{39th International Cosmic Ray Conference (ICRC2025)\\
 15–24 July 2025\\
Geneva, Switzerland\\}

\begin{document}
\maketitle

\section{Introduction}\label{sec:intro}
    \paragraph*{}Cosmic rays~(CRs) with energies between $\sim10^{11.5}$ and $10^{16.5}~\mathrm{eV}$ per nucleon span the region of the all-particle spectrum where the so-called knee appears, and where the transition from Galactic to potentially extragalactic accelerators is expected. Accurate measurements of the total flux and elemental composition within this energy range are therefore pivotal for discriminating among competing models of CR acceleration and propagation. Ground-based observatories, such as the High-Altitude Water Cherenkov~(HAWC) Observatory, record the footprints of extensive air showers~(EAS) produced when primary particles interact with the atmosphere. The collected information, time and charge measurements, can be translated into estimations of the main EAS parameters, the most relevant of which are the primary energy, the arrival direction, and core position.
    \paragraph*{}HAWC is designed to detect high-energy gamma rays and cosmic rays. It is located on the slope of the Sierra Negra volcano in Puebla, Mexico, at an elevation of 4100 m a.s.l. HAWC’s elevation allows for efficient detection of the secondary particles produced before they are significantly absorbed by the atmosphere. The array comprises 300 water Cherenkov detectors (WCDs), each 7.3 m in diameter and 4.5 m in height, covering an area of 22,000~$m^2$. Each tank is instrumented with four upward-facing photomultiplier tubes (PMTs), which capture the Cherenkov light produced by the secondary particles of the EAS traveling through the water~\cite{Abeysekara2013}. HAWC is sensitive to showers induced by cosmic rays ranging from TeV to a few PeV energies~\cite{Abeysekara2017CR}. The extensive data set collected by the observatory allows us to develop advanced reconstruction methods, including machine learning algorithms, to improve the measurements of the shower parameters, such as the energy estimator for gamma rays~\cite{crab2019} and cosmic rays~\cite{Alvarado2023}.
    \paragraph*{}The first cosmic-ray energy estimator implemented in HAWC is a traditional model that chooses the maximum likelihood match between the detected shower and four-dimensional probability tables\footnote{The dimensions mentioned are the zenith angle, primary energy, distance from the HAWC central array to the shower core position, and PMT signal amplitude.}~\cite{Abeysekara2017CR}. The first attempt to provide an alternative cosmic-ray energy estimator using machine learning techniques was reported at the previous ICRC~\cite{Alvarado2023}, with the objective of obtaining a more accurate prediction of this observable. In this contribution, we present an update on the application of a Multilayer Perceptron (MLP) for predicting cosmic-ray energy. We report three models trained with TensorFlow, which allows the development of more complex architectures and other characteristics (more details are provided in Section~\ref{sec:mlp}). These models were trained and evaluated using HAWC simulation data. To ensure a robust assessment, the models were also applied to one day of HAWC data to obtain the cosmic-ray spectrum. The comparative analysis between the machine learning models and the official cosmic-ray energy estimator is provided in Section~\ref{sec:aandr}. Finally, in Section~\ref{sec:disandcon}, we summarize this contribution.

\section{Data Set}\label{sec:dataset}
    \paragraph*{}Two types of data were used in this contribution: simulation and real data, both of which were produced following the HAWC official procedure~\cite{crab2017}. Eight cosmic-ray particles (hydrogen, helium, carbon, oxygen, neon, magnesium, silicon, and iron) were simulated with energies between $\sim$5~GeV and 2~PeV, following a power-law spectrum with an index of -2. The hadronic model specified in CORSIKA is FLUKA and QGSJET-II-04 for low and high energies, respectively, with zenith angles ranging from $0^\circ$ to $60^\circ$~\cite{ALFARO2025103077}. The number of remaining events after the reconstruction process is 14 million\footnote{Half of the total events are protons.}. On the other hand, one complete day of HAWC data was used to perform a robust evaluation of the MLP models.
\section{Multilayer Perceptron (MLP)}\label{sec:mlp}
    \paragraph*{}The machine learning model used in this contribution is the MLP. Three models were implemented using the same architecture and training specifications; the difference lies in how the input information is provided. For these models, there is a common set of input features: the total charge collected by the detectors within nine concentric annuli around the shower core~\cite{crab2017}; the amplitude of the core fit to a simplified Nishimura–Kamata–Greisen~(NKG) function~\cite{crab2019}; and the perceptual area activated during the event. The differing input information includes the fraction of PMTs activated ($f_{hit}$), the arrival direction, and the core position. In these cases, feature transformations are applied to vary the input representations across the models (see Table~\ref{tab:features}). The input parameters of each model were normalized using the $z$‑score method, which rescales each feature to have zero mean and unit variance~\cite{James2021}.

\begin{table}
    \centering
    \caption{The combination of the input information that differs between the MLPs is achieved using feature transformations. The raw information includes the number of PMTs activated during the event (nHit), the number of PMTs available during the detection (nChavail), the zenith angle in radians, and the X and Y core positions. The feature transformations used are: the fraction of PMTs activated ($f_{hit}$), the cosine of the zenith angle, and the distance between the core position and the center of the HAWC array ($R_{core}$).}
    \label{tab:features}
    \begin{tabular}{|c|c|c|}
        \hline
        \textbf{MLP-V1} & \textbf{MLP-V2} & \textbf{MLP-V3} \\
        \hline
        \hline
        \multirow{2}{2em}{$f_{hit}$} & nHit   & \multirow{2}{2em}{$f_{hit}$}\\
        & nChavail &\\
        \hline
        cos(zenith angle) & zenith angle & zenith angle\\
        \hline
         \multirow{2}{3em}{$R_{core}$}& core X & core X\\
         & core Y & core Y\\
         \hline
    \end{tabular}
\end{table}

    \paragraph*{} The architecture implemented is $x:256:128:64:32:1$, where $x$ is the number of input parameters; there are four hidden dense layers with ReLU activation functions and one linear output layer that provides an estimation of the energy based on the input information. During the training stage, Mean Squared Error (MSE) was used as the loss function, Stochastic Gradient Descent as the optimization algorithm, a learning rate of 0.01, and a maximum of 15 epochs with batches of 64. The data was split into three groups: 64\%, 16\%, and 20\% for training, validation, and evaluation, respectively. Events from eight species were used, selecting those that passed successful reconstruction of the core and arrival direction, with a minimum fraction of 20\% of PMTs with signal relative to the total array ($f_{hit} > 0.2$), zenith angles from $0^\circ$ to $35^\circ$, and at least 40 PMTs with signal within a radius of 40 meters from the shower core~\cite{ALFARO2025103077}.

\begin{figure}[!ht]
    \begin{subfigure}{0.49\textwidth}
         \includegraphics[width=\textwidth]{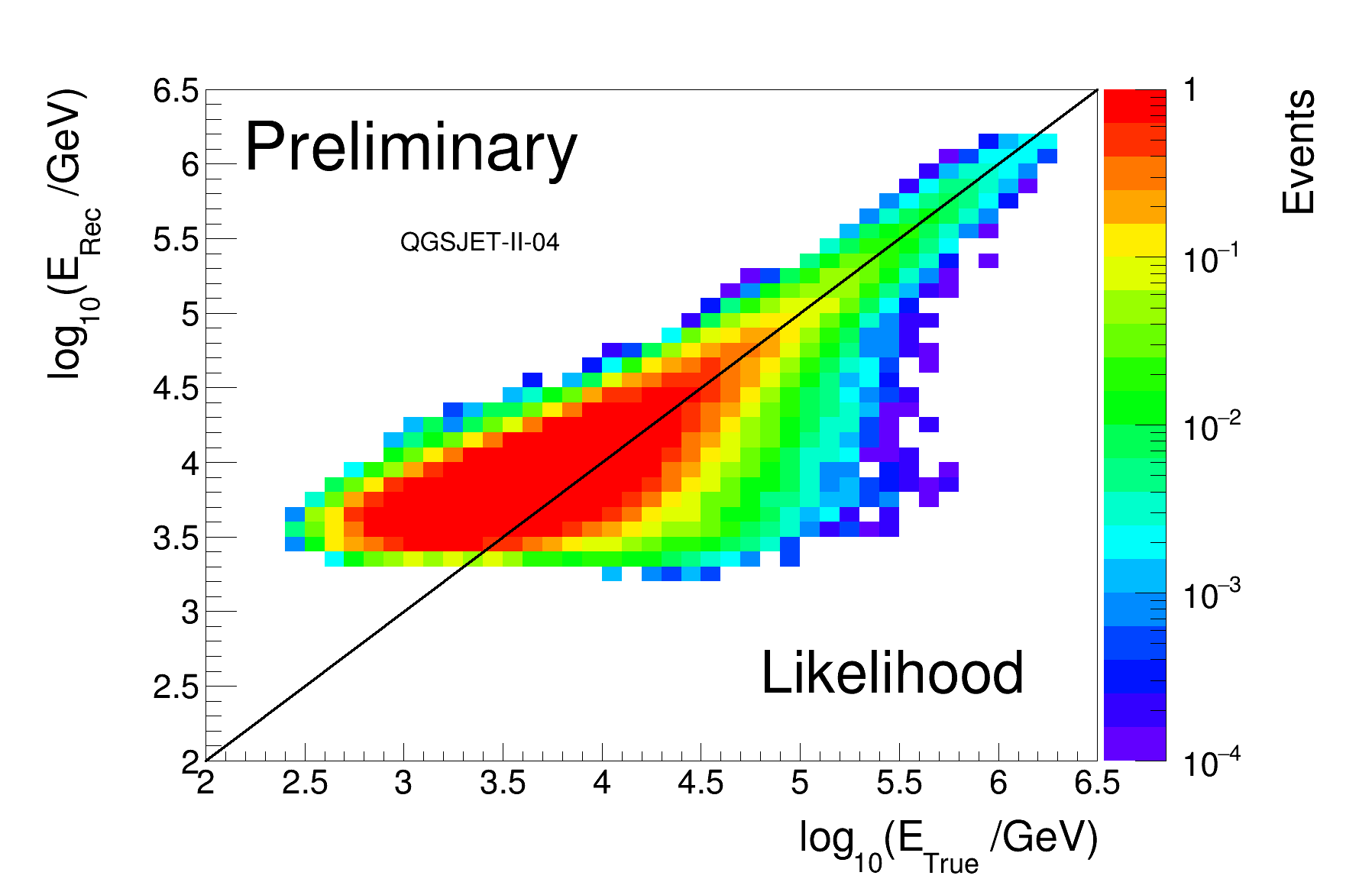}
         \caption{Likelihood.}
         \label{fig:h2dlh}
     \end{subfigure}
     \hfill
     \begin{subfigure}{0.49\textwidth}
         \includegraphics[width=\textwidth]{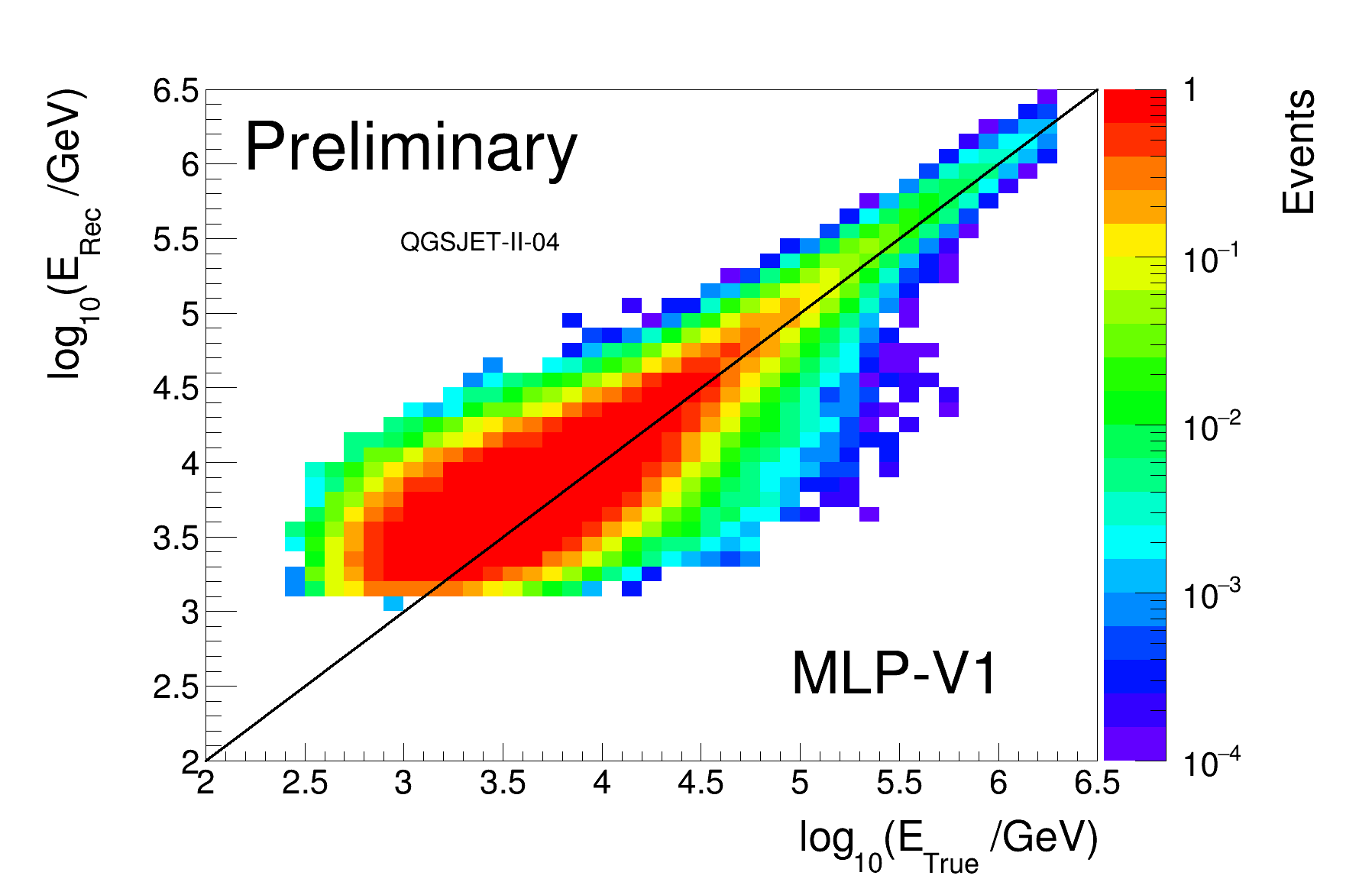}
         \caption{MLP-V1.}
         \label{fig:h2dmlpv1}
     \end{subfigure}
     \medskip
     \begin{subfigure}{0.49\textwidth}
         \includegraphics[width=\textwidth]{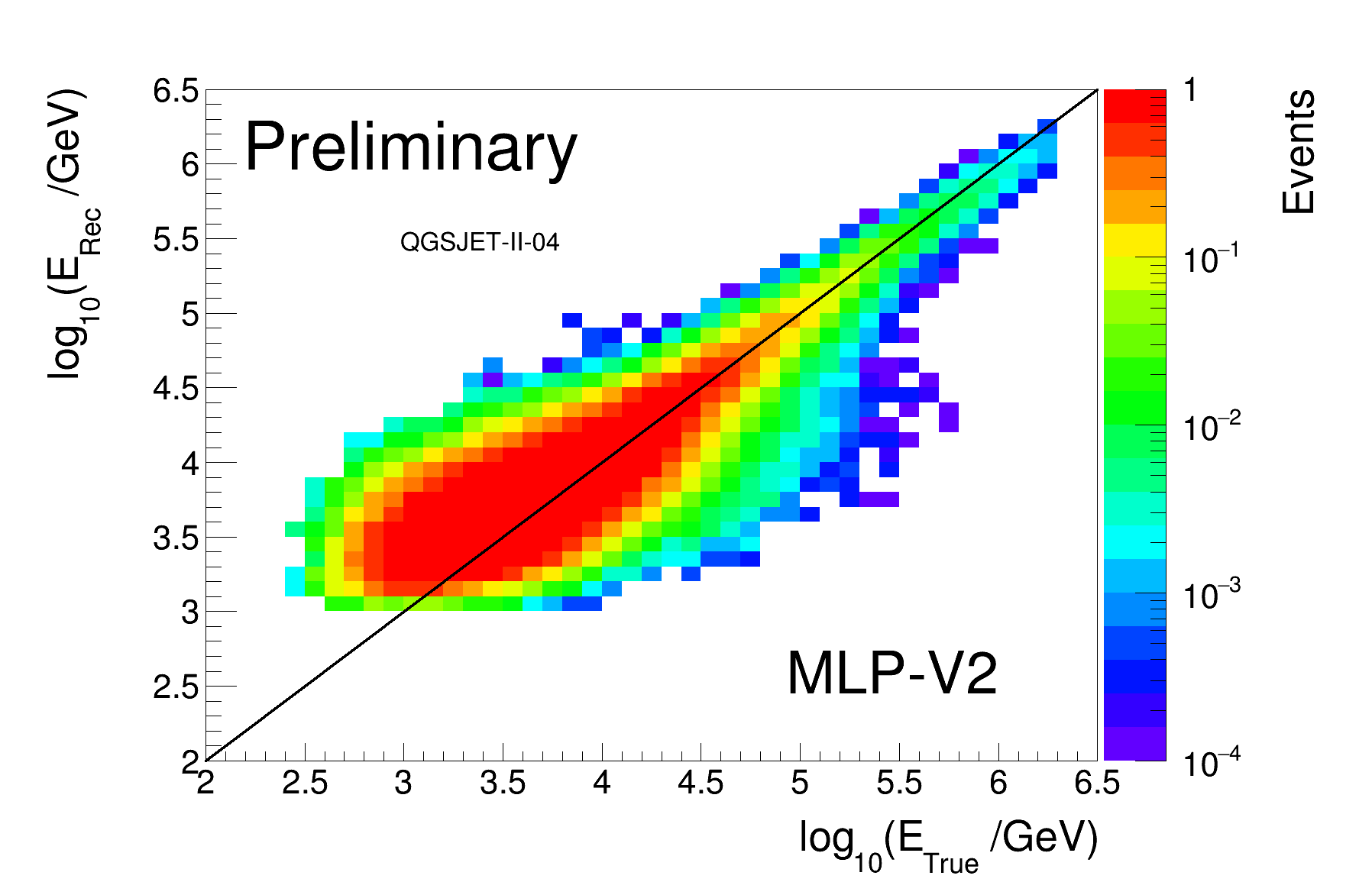}
         \caption{MLP-V2.}
         \label{fig:h2dmlpv2}
     \end{subfigure}
     \hfill
     \begin{subfigure}{0.49\textwidth}
         \includegraphics[width=\textwidth]{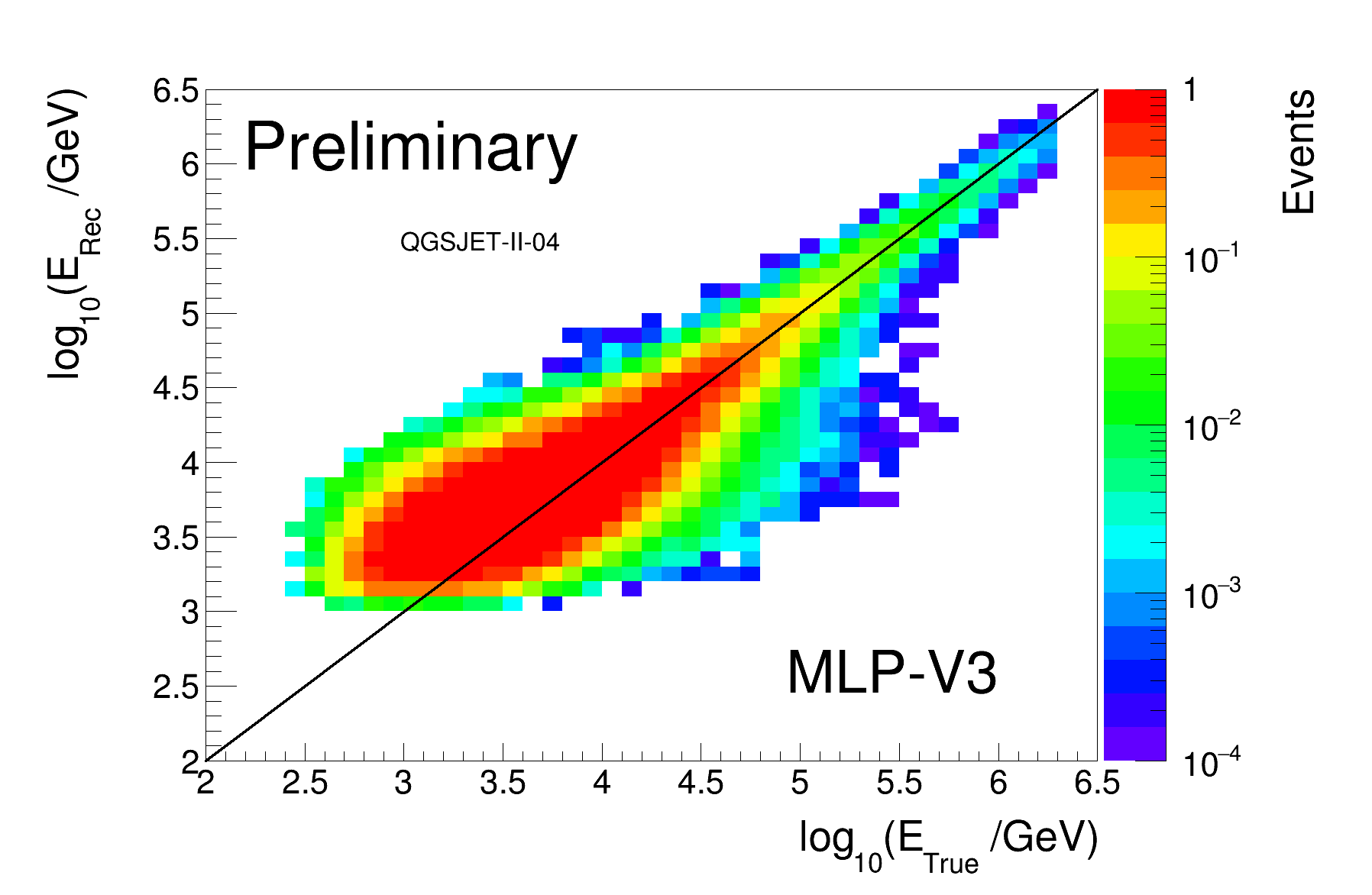}
         \caption{MLP-V3.}
         \label{fig:h2dmlpv3}
     \end{subfigure} 
     \caption{Density maps of reconstructed energy versus true energy are shown using proton showers for the Likelihood method and the three MLP models proposed in this contribution. The black line indicates the region where the reconstructed and true values are equal ($E_{\mathrm{rec}} = E_{\mathrm{true}}$), representing the ideal reconstruction.}
     \label{fig:h2dmodels}
\end{figure}
    
\section{Analysis and Results}\label{sec:aandr}
\begin{figure}[!ht]
    \centering
    \begin{subfigure}{0.8\textwidth}
         \includegraphics[width=\textwidth]{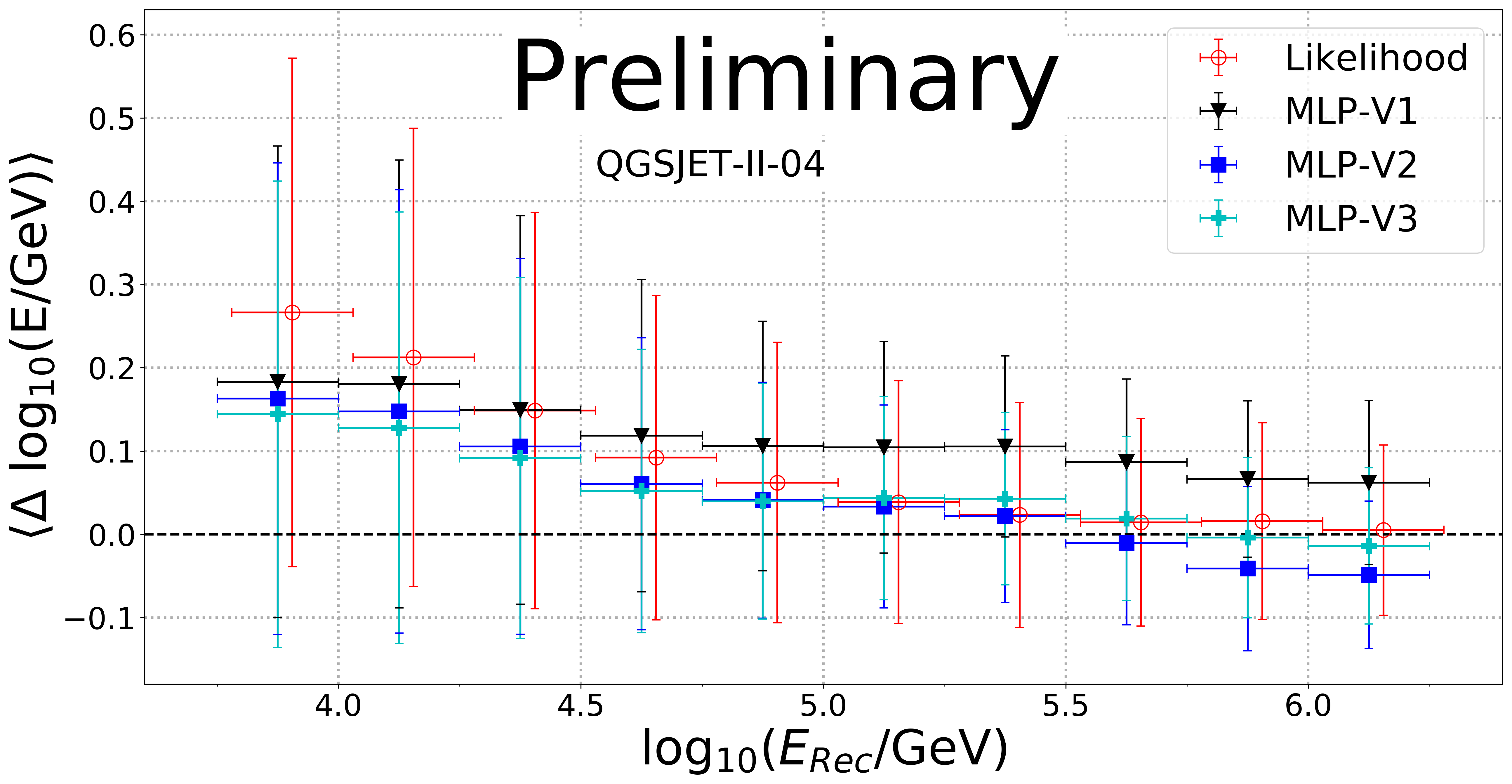}
         \caption{Mean bias.}
         \label{fig:bias}
     \end{subfigure}
     \medskip
     \begin{subfigure}{0.8\textwidth}
         \includegraphics[width=\textwidth]{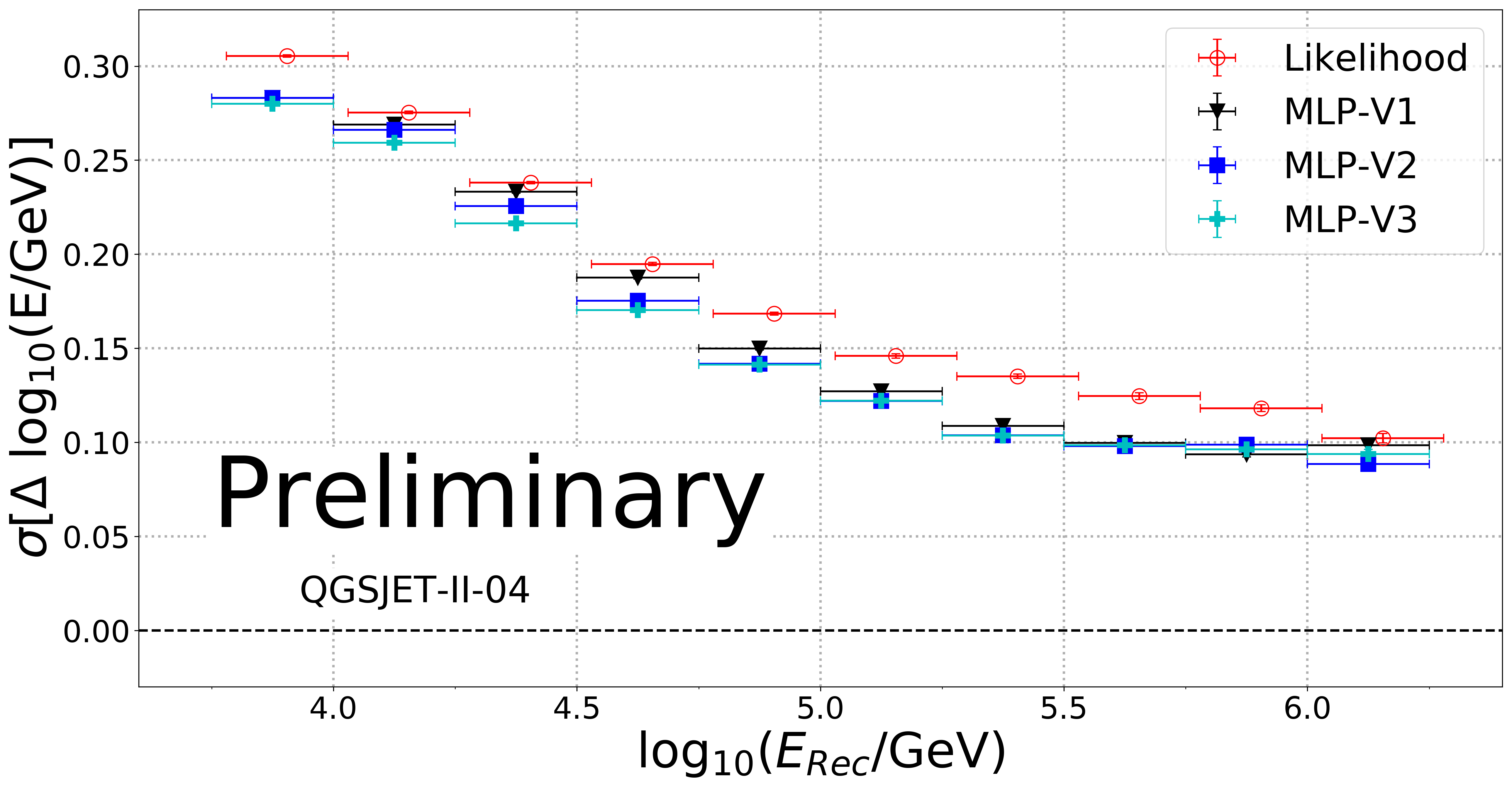}
         \caption{Energy resolution.}
         \label{fig:reso}
     \end{subfigure}
     \caption{The performance of the Likelihood method and the three new models is presented. In panel (a), the model fidelity is shown; this represents how close the predictions are to the true values. It is quantified by the mean of the bias computed for each energy bin, $\langle\Delta\log_{10}(\text{E/GeV})\rangle$, where $\Delta\log_{10}(\text{E/GeV})=\log_{10}(\text{E}_{\text{rec}})-\log_{10}(\text{E}_{\text{true}})$. In panel (b), the energy resolution is shown, defined as $\sigma [\Delta \log_{10}(\text{E/GeV})]$.}
     \label{fig:performance}
\end{figure}

    \paragraph*{}As mentioned in the previous section, the MLPs were trained using simulations of eight species to provide a comprehensive overview of the behavior of these particles. However, in the evaluation phase, only proton events are reported. Figure~\ref{fig:h2dmodels} shows the correlation between reconstructed energy and true energy for the Likelihood method and the three MLP models. In HAWC, the official cosmic-ray energy estimator is the Likelihood method. However, due to a recent software update aimed at improving gamma-ray source analysis, the performance of the Likelihood estimator has decreased. This can be seen in Figure~\ref{fig:h2dlh}, where there is a region in which the event prediction is almost one order of magnitude lower than the true value. On the other hand, the MLPs address this inconsistency by reducing this region and pushing those events closer to the identity line (Figures~\ref{fig:h2dmlpv1}, \ref{fig:h2dmlpv2}, and \ref{fig:h2dmlpv3} correspond to the MLP-V1, MLP-V2, and MLP-V3 models, respectively). These observations are supported by Figure~\ref{fig:performance}, where the mean bias ($\langle\Delta\log_{10}(\text{E/GeV})\rangle$, top panel) and energy resolution ($\sigma[\Delta\log_{10}(\text{E/GeV})]$, bottom panel) are computed in energy bins. The MLP models show a lower mean bias value than the Likelihood estimator, meaning they provide estimates closer to the true value at lower energies. Furthermore, the models group the events closer to the ideal reconstruction line, as reflected in the bottom panel, where the energy resolution is improved across all energy bins.
        
\begin{figure}[!ht]
    \centering
    \begin{subfigure}{0.8\textwidth}
         \includegraphics[width=\textwidth]{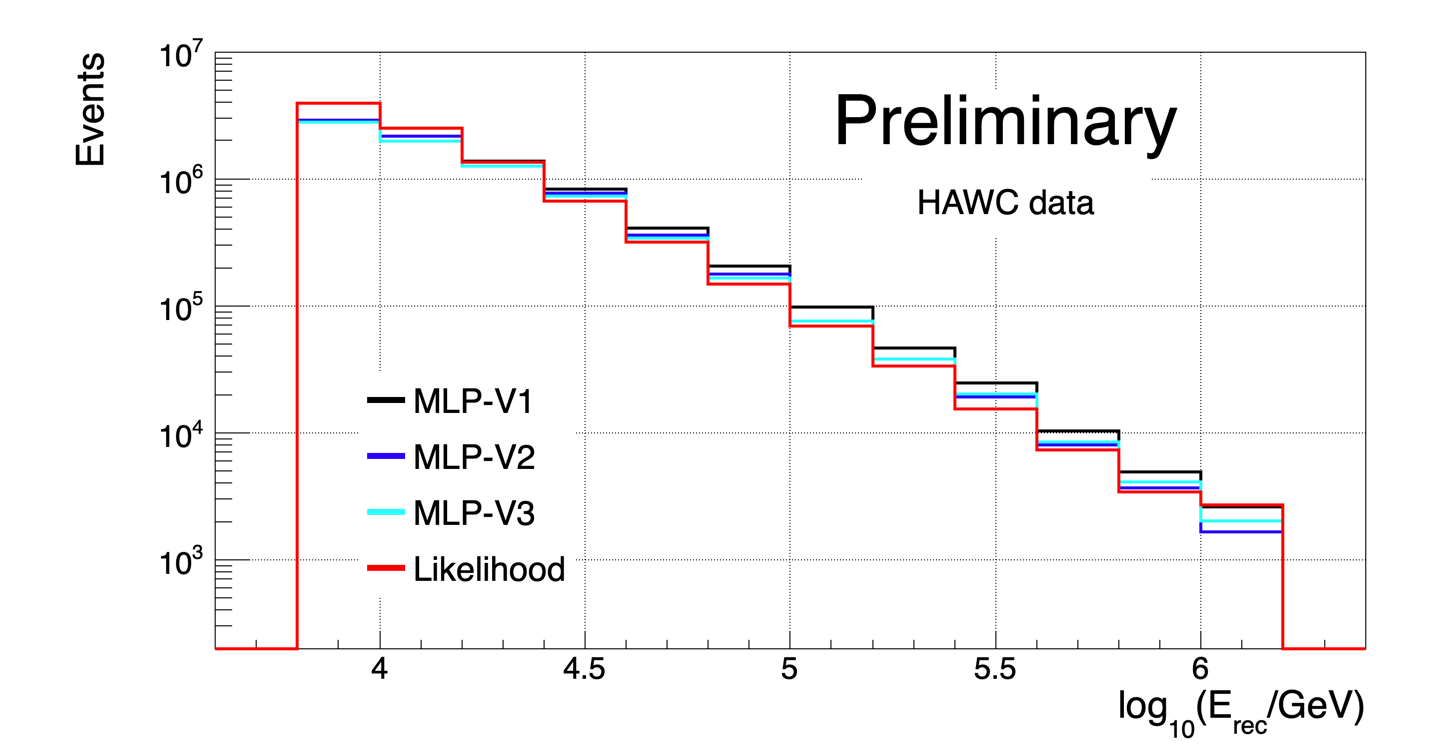}
         \caption{Histogram of the model reconstruction using one day of HAWC data.}
         \label{fig:histord}
     \end{subfigure}
     \medskip
     \begin{subfigure}{0.8\textwidth}
         \includegraphics[width=\textwidth]{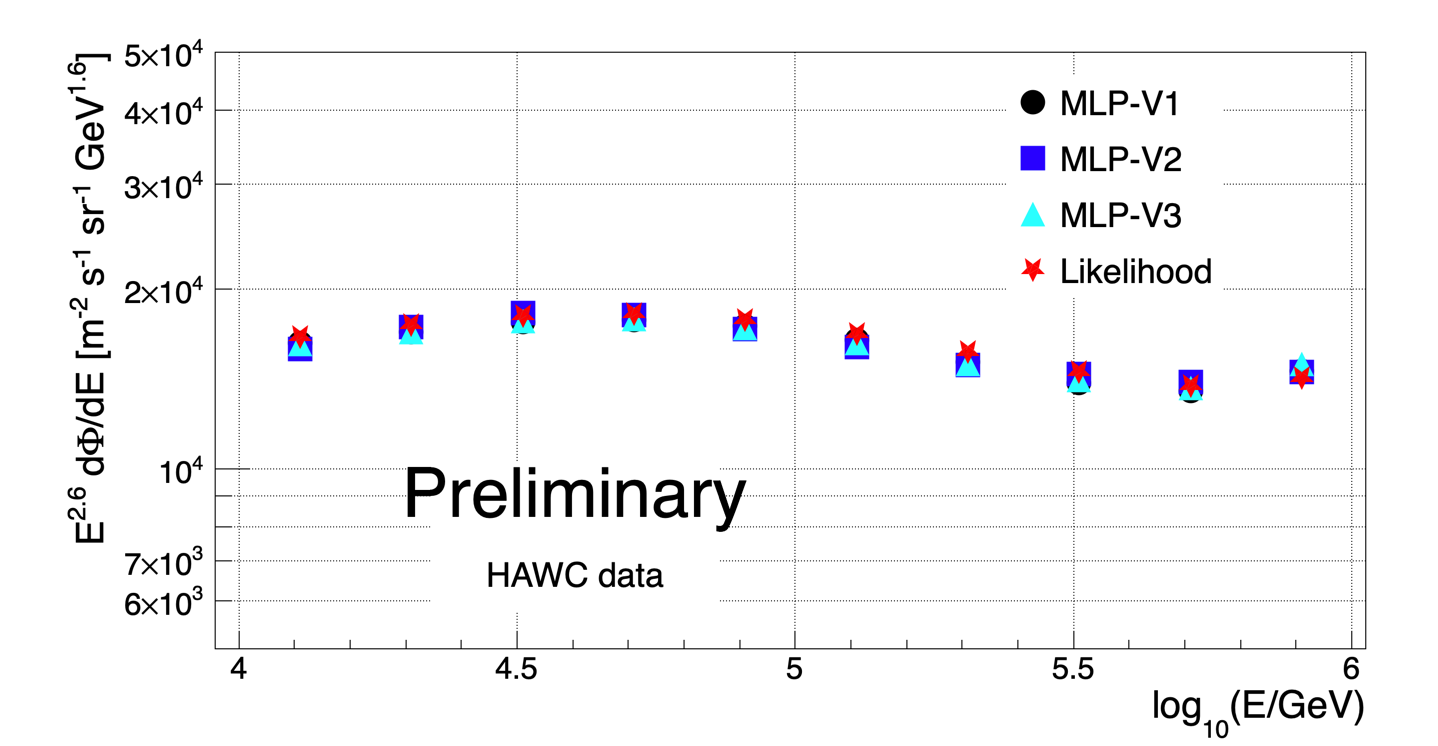}
         \caption{Spectrum of the model reconstruction using one day of HAWC data.}
         \label{fig:spectrumrd}
     \end{subfigure}
     \caption{Performance of the models when applied to one day of HAWC real data. In (a), the distributions of the model outputs are shown; these models exhibit almost identical behavior. On the other hand, in (b), the one-day cosmic-ray spectrum obtained using the unfolding technique (as described in~\cite{ALFARO2025103077}) is presented. In summary, all models report the same behavior, indicating that the new MLP models are consistent with the Likelihood method. This consistency suggests that the trained models are robust and reliable.}
     \label{fig:realdata}
\end{figure}

    \paragraph*{}The robust evaluation consists of testing the models with real data. For faster comparison between models, one day of HAWC real data is used to reconstruct the cosmic-ray spectrum. The spectrum is obtained using the Unfolding method, following the procedure described in~\cite{ALFARO2025103077,Abeysekara2017CR}. In this contribution, we report the output model distributions (Figure~\ref{fig:histord}) and the unfolded spectrum (Figure~\ref{fig:spectrumrd}). The distributions show consistency across all models, with small fluctuations in some energy ranges. On the other hand, the spectra from the MLP models are in good agreement with the official energy reconstructor, Likelihood method. These models show small fluctuations similar to those observed in the histogram distributions.

\section{Discussion and conclusion}\label{sec:disandcon}
    \paragraph*{}Supervised machine learning algorithms have improved in recent years. They have demonstrated the ability to predict values based on input information and, in many cases, provide better results than classical techniques. In this contribution, the MLP model, using  TensorFlow package for training, was explored to address a regression task at the HAWC observatory; this task is to estimate the energy of the detected cosmic-ray events. Three MLPs were trained using simulations of eight particle types, with the same training configuration but varying the input feature transformations. These models were compared to the official cosmic-ray energy estimator, Likelihood, using both simulated and real HAWC data. According to the simulation results, the three MLPs achieved a lower mean bias at lower energies, indicating better accuracy in event energy estimation and improved energy resolution, which suggests greater confidence in the predictions (see Figures~\ref{fig:h2dmodels} and \ref{fig:performance}).
    \paragraph*{}A robust evaluation was performed by computing the cosmic-ray spectrum using one day of HAWC real data. This process makes it possible to determine whether the models were properly trained, whether the MLPs simply memorized the events or began to overfit or underfit the data. Figure~\ref{fig:realdata} shows the output distributions of the Likelihood and the three MLP models, along with their corresponding spectra computed using the unfolding technique. From both figures, it can be concluded that the MLP models produce consistent results, as they follow the same behavior as the Likelihood method. These results suggest that the MLP models provide a reliable alternative for estimating the energy of cosmic-ray-induced showers. Moreover, the improved performance of the MLPs (reflected in lower estimation errors, better mean bias, and higher confidence in the energy predictions, better energy resolution) makes them a promising option for future updates of the energy reconstruction method.
    
\acknowledgments{
We acknowledge the support from: the US National Science Foundation (NSF); the US Department of Energy Office of High-Energy Physics; the Laboratory Directed Research and Development (LDRD) program of Los Alamos National Laboratory; Consejo Nacional de Ciencia y Tecnolog\'{i}a (CONACyT), M\'{e}xico, grants LNC-2023-117, 271051, 232656, 260378, 179588, 254964, 258865, 243290, 132197, A1-S-46288, A1-S-22784, CF-2023-I-645, CBF2023-2024-1630, c\'{a}tedras 873, 1563, 341, 323, Red HAWC, M\'{e}xico; DGAPA-UNAM grants IG101323, IN111716-3, IN111419, IA102019, IN106521, IN114924, IN110521 , IN102223; VIEP-BUAP; PIFI 2012, 2013, PROFOCIE 2014, 2015; the University of Wisconsin Alumni Research Foundation; the Institute of Geophysics, Planetary Physics, and Signatures at Los Alamos National Laboratory; Polish Science Centre grant, 2024/53/B/ST9/02671; Coordinaci\'{o}n de la Investigaci\'{o}n Cient\'{i}fica de la Universidad Michoacana; Royal Society - Newton Advanced Fellowship 180385; Gobierno de España and European Union-NextGenerationEU, grant CNS2023- 144099; The Program Management Unit for Human Resources \& Institutional Development, Research and Innovation, NXPO (grant number B16F630069); Coordinaci\'{o}n General Acad\'{e}mica e Innovaci\'{o}n (CGAI-UdeG), PRODEP-SEP UDG-CA-499; Institute of Cosmic Ray Research (ICRR), University of Tokyo. H.F. acknowledges support by NASA under award number 80GSFC21M0002. C.R. acknowledges support from National Research Foundation of Korea (RS-2023-00280210). We also acknowledge the significant contributions over many years of Stefan Westerhoff, Gaurang Yodh and Arnulfo Zepeda Dom\'inguez, all deceased members of the HAWC collaboration. Thanks to Scott Delay, Luciano D\'{i}az and Eduardo Murrieta for technical support.}

\bibliographystyle{JHEP} 
\bibliography{biblio}
\clearpage
\section*{Full Author List: \ HAWC Collaboration}
\scriptsize
\noindent
%
\vskip2cm
\noindent

R. Alfaro$^{1}$,
C. Alvarez$^{2}$,
A. Andrés$^{3}$,
E. Anita-Rangel$^{3}$,
M. Araya$^{4}$,
J.C. Arteaga-Velázquez$^{5}$,
D. Avila Rojas$^{3}$,
H.A. Ayala Solares$^{6}$,
R. Babu$^{7}$,
P. Bangale$^{8}$,
E. Belmont-Moreno$^{1}$,
A. Bernal$^{3}$,
K.S. Caballero-Mora$^{2}$,
T. Capistrán$^{9}$,
A. Carramiñana$^{10}$,
F. Carreón$^{3}$,
S. Casanova$^{11}$,
S. Coutiño de León$^{12}$,
E. De la Fuente$^{13}$,
D. Depaoli$^{14}$,
P. Desiati$^{12}$,
N. Di Lalla$^{15}$,
R. Diaz Hernandez$^{10}$,
B.L. Dingus$^{16}$,
M.A. DuVernois$^{12}$,
J.C. Díaz-Vélez$^{12}$,
K. Engel$^{17}$,
T. Ergin$^{7}$,
C. Espinoza$^{1}$,
K. Fang$^{12}$,
N. Fraija$^{3}$,
S. Fraija$^{3}$,
J.A. García-González$^{18}$,
F. Garfias$^{3}$,
N. Ghosh$^{19}$,
A. Gonzalez Muñoz$^{1}$,
M.M. González$^{3}$,
J.A. Goodman$^{17}$,
S. Groetsch$^{19}$,
J. Gyeong$^{20}$,
J.P. Harding$^{16}$,
S. Hernández-Cadena$^{21}$,
I. Herzog$^{7}$,
D. Huang$^{17}$,
P. Hüntemeyer$^{19}$,
A. Iriarte$^{3}$,
S. Kaufmann$^{22}$,
D. Kieda$^{23}$,
K. Leavitt$^{19}$,
H. León Vargas$^{1}$,
J.T. Linnemann$^{7}$,
A.L. Longinotti$^{3}$,
G. Luis-Raya$^{22}$,
K. Malone$^{16}$,
O. Martinez$^{24}$,
J. Martínez-Castro$^{25}$,
H. Martínez-Huerta$^{30}$,
J.A. Matthews$^{26}$,
P. Miranda-Romagnoli$^{27}$,
P.E. Mirón-Enriquez$^{3}$,
J.A. Montes$^{3}$,
J.A. Morales-Soto$^{5}$,
M. Mostafá$^{8}$,
M. Najafi$^{19}$,
L. Nellen$^{28}$,
M.U. Nisa$^{7}$,
N. Omodei$^{15}$,
E. Ponce$^{24}$,
Y. Pérez Araujo$^{1}$,
E.G. Pérez-Pérez$^{22}$,
Q. Remy$^{14}$,
C.D. Rho$^{20}$,
D. Rosa-González$^{10}$,
M. Roth$^{16}$,
H. Salazar$^{24}$,
D. Salazar-Gallegos$^{7}$,
A. Sandoval$^{1}$,
M. Schneider$^{1}$,
G. Schwefer$^{14}$,
J. Serna-Franco$^{1}$,
A.J. Smith$^{17}$
Y. Son$^{29}$,
R.W. Springer$^{23}$,
O. Tibolla$^{22}$,
K. Tollefson$^{7}$,
I. Torres$^{10}$,
R. Torres-Escobedo$^{21}$,
R. Turner$^{19}$,
E. Varela$^{24}$,
L. Villaseñor$^{24}$,
X. Wang$^{19}$,
Z. Wang$^{17}$,
I.J. Watson$^{29}$,
H. Wu$^{12}$,
S. Yu$^{6}$,
S. Yun-Cárcamo$^{17}$,
H. Zhou$^{21}$,

\vskip2cm
\noindent

$^{1}$Instituto de F\'{i}sica, Universidad Nacional Autónoma de México, Ciudad de Mexico, Mexico,
$^{2}$Universidad Autónoma de Chiapas, Tuxtla Gutiérrez, Chiapas, México,
$^{3}$Instituto de Astronom\'{i}a, Universidad Nacional Autónoma de México, Ciudad de Mexico, Mexico,
$^{4}$Universidad de Costa Rica, San José 2060, Costa Rica,
$^{5}$Universidad Michoacana de San Nicolás de Hidalgo, Morelia, Mexico,
$^{6}$Department of Physics, Pennsylvania State University, University Park, PA, USA,
$^{7}$Department of Physics and Astronomy, Michigan State University, East Lansing, MI, USA,
$^{8}$Temple University, Department of Physics, 1925 N. 12th Street, Philadelphia, PA 19122, USA,
$^{9}$Universit\`a degli Studi di Torino, I-10125 Torino, Italy,
$^{10}$Instituto Nacional de Astrof\'{i}sica, Óptica y Electrónica, Puebla, Mexico,
$^{11}$Institute of Nuclear Physics Polish Academy of Sciences, PL-31342 11, Krakow, Poland,
$^{12}$Dept. of Physics and Wisconsin IceCube Particle Astrophysics Center, University of Wisconsin{\textemdash}Madison, Madison, WI, USA,
$^{13}$Departamento de F\'{i}sica, Centro Universitario de Ciencias Exactase Ingenierias, Universidad de Guadalajara, Guadalajara, Mexico, 
$^{14}$Max-Planck Institute for Nuclear Physics, 69117 Heidelberg, Germany,
$^{15}$Department of Physics, Stanford University: Stanford, CA 94305–4060, USA,
$^{16}$Los Alamos National Laboratory, Los Alamos, NM, USA,
$^{17}$Department of Physics, University of Maryland, College Park, MD, USA,
$^{18}$Tecnologico de Monterrey, Escuela de Ingenier\'{i}a y Ciencias, Ave. Eugenio Garza Sada 2501, Monterrey, N.L., Mexico, 64849,
$^{19}$Department of Physics, Michigan Technological University, Houghton, MI, USA,
$^{20}$Department of Physics, Sungkyunkwan University, Suwon 16419, South Korea,
$^{21}$Tsung-Dao Lee Institute \& School of Physics and Astronomy, Shanghai Jiao Tong University, 800 Dongchuan Rd, Shanghai, SH 200240, China,
$^{22}$Universidad Politecnica de Pachuca, Pachuca, Hgo, Mexico,
$^{23}$Department of Physics and Astronomy, University of Utah, Salt Lake City, UT, USA, 
$^{24}$Facultad de Ciencias F\'{i}sico Matemáticas, Benemérita Universidad Autónoma de Puebla, Puebla, Mexico, 
$^{25}$Centro de Investigaci\'on en Computaci\'on, Instituto Polit\'ecnico Nacional, M\'exico City, M\'exico,
$^{26}$Dept of Physics and Astronomy, University of New Mexico, Albuquerque, NM, USA,
$^{27}$Universidad Autónoma del Estado de Hidalgo, Pachuca, Mexico,
$^{28}$Instituto de Ciencias Nucleares, Universidad Nacional Autónoma de Mexico, Ciudad de Mexico, Mexico, 
$^{29}$University of Seoul, Seoul, Rep. of Korea,
$^{30}$Departamento de Física y Matemáticas, Universidad de Monterrey, Av.~Morones Prieto 4500, 66238, San Pedro Garza García NL, México

\end{document}